\documentclass[useAMS,usenatbib,usegraphicx]{mn2e}
\usepackage{aas_macros,amsmath,amssymb}

\usepackage[dvips]{color}

\title[Imaging the asymmetric dust shell around CI Cam]{Imaging the asymmetric dust shell around CI Cam with long baseline optical interferometry}

\author[Thureau et al.]{N.D. Thureau$^{1}$\thanks{E-mail:nt15@st-andrews.ac.uk}, J.D. Monnier$^{2}$, W.A. Traub$^{3}$, R. Millan-Gabet$^{4}$, E. Pedretti$^{1}$, 
  \newauthor 
J.-P. Berger$^{5}$, M.R. Garcia$^{6}$, F.P. Schloerb$^{7}$, A.-K. Tannirkulam$^{2}$\\
$^{1}$ School of  Physics and Astronomy, University of  St Andrews, North Haugh, St Andrews KY16 9SS, Scotland\\
$^{2}$ University of Michigan, Astronomy dept., 914 Dennison bldg., 500 Church street, Ann Arbor, MI, 40109, USA\\\
$^{3}$ Jet Propulsion Laboratory, M/S 301--451, 4800 Oak Grove Drive, Pasadena, CA, 91109, USA\\
$^{4}$ Michelson Science Center, California Institute of Technology, 770 S. Wilson Ave. MS 100--22, Pasadena, CA 91125, USA\\
$^{5}$ Laboratoire d'Astrophysique de l'Observatoire de Grenoble (LAOG), 414 rue de la Piscine, BP 53--X, Grenoble, France\\
$^{6}$ Smithsonian Astrophysical Observatory, Center for Astrophysics, 60 Garden Street, Cambridge  MA 02138, USA\\
$^{7}$ University of Massachusetts, Department of Astronomy, Amherst, MA, 01003--4610, USA
}

\begin{document}

\date{Accepted 2009 April 20.  Received 2009 April 16; in original form 2008 October 7}

\pagerange{\pageref{firstpage}--\pageref{lastpage}} \pubyear{xxxx}

\maketitle

\label{firstpage}

\begin{abstract}
We present the first high angular resolution observation of the B[e] star/X-ray transient object CI Cam, performed with the two-telescope Infrared Optical Telescope Array (IOTA), its upgraded three-telescope version (IOTA3T) and the Palomar Testbed Interferometer (PTI). Visibilities and closure phases were obtained using the IONIC-3 integrated optics beam combiner. CI Cam was observed in the near-infrared H and K spectral bands, wavelengths well suited to measure the size and study the geometry of the hot dust surrounding CI Cam.
The analysis of the visibility data over an 8 year period from soon after the 1998 outburst to 2006 shows that the dust visibility has not changed over the years. 
The visibility data shows that CI Cam is elongated which confirms the disc-shape of the circumstellar environment and totally rules out the hypothesis of a spherical dust shell.
Closure phase measurements show direct evidence of asymmetries in the circumstellar environment of CI Cam and we conclude that the dust surrounding CI Cam lies in an inhomogeneous disc seen at an angle. The near-infrared dust emission appears as an elliptical skewed Gaussian ring with a major axis a = $7.58 \pm 0.24$ mas, an axis ratio r = $0.39 \pm 0.03$ and a position angle $\theta=35 \pm 2 ^{\circ}$.
\end{abstract}


\section{Introduction}
The star CI Cam (MWC 84, IRAS 04156+5552), optical counterpart of XTE J0421+560, has been intensively observed since its 1998 outburst.  \citet{2002A&A...392..991H} and \citet{2002ApJ...565.1169R} classified it as a high luminosity ($> 10^{4}~L_{\sun}$) B[e] supergiant of B0-B2 spectral type with an invisible compact companion (neutron star or black hole) and estimate the distance to CI Cam to be $\gtrsim$5~kpc. \citet{2006ARep...50..664B} classified it as a B4III-V with a white dwarf companion itself surrounded by an accretion disc at a distance d=1.1-1.9~kpc. The distance to CI Cam remains uncertain and values from the literature range from 0.2 to 17~kpc with 5~kpc being the value generally adopted. The primary component of CI Cam displays all the observational characteristics of a B[e] star: strong Balmer emission lines, optical forbidden lines and a strong infrared excess. The infrared excess was first reported by \citet{1973MNRAS.161..145A} who found that CI Cam H-K and K-L large colour indices showed evidence of re-radiation from circumstellar dust. 

The 1998 x-ray outburst of CI Cam was very unusual in that it was extremely fast and bright - rising to a peak luminosity of $3 \times 10^{38}$ erg/sec (assuming 5kpc distance) within 12 hours and decaying with an e-folding time of 0.6 days \citep{1998IAUC.6855....1S,1999AstL...25..294R}. While this peak luminosity is typical of that reached by the classical black hole x-ray novae \citep{2006csxs.book..157M}, the outburst is much faster than the week long rise and  months long decay typical of these systems.  On the other hand, the 1998 outburst was not quite as fast as those typical of the supergiant fast x-ray transients \citep{2006xru..conf..165N,2006ApJ...638..974S} which contain OB supergiant stars, but these typically only reach $10^{36}$ erg/sec. One object which did have a similarly short and bright outburst is V4641 Sgr, which does contain a black hole, but the other member of this binary is a B9III \citep{2001ApJ...555..489O}, not a supergiant B[e] as in CI Cam. The high x-ray luminosity and simultaneous radio eruption have led to the suggestion that the CI Cam system containse a black hole \citep{2002ApJ...565.1169R,1999ApJ...527..345B}.

CI Cam circumstellar gaseous environment has been subsequently probed using spectroscopy and spectropolarimetry techniques. \citet{2002A&A...392..991H} concluded from a series of spectroscopic observations that the dust surrounding CI Cam cannot exist in a spherical circumstellar shell and can only form in the outer region of its equatorial gas disc. As such, the dusty environment surrounding CI Cam is expected to be ring shaped. From high resolution spectroscopy observations taken 3 years after the outburst \citet{2002A&A...390..627M} showed that the emission lines are clearly resolved and display a triple-peaked structure suggesting an intermediate inclination angle of the circumstellar gas disc which has been recently confirmed by \citet{2007AJ....133.1478Y}. One could expect a polarisation signature of the disc orientation but spectropolarimetric observations results \citep{2000A&A...355..256I} remain inconclusive. 

CI Cam's dust region, the source of the infrared emission, has been discussed by  \citet{1995A&AT....6..251M},   \citet{1999ApJ...527..345B} and \citet{2000A&A...356...50C} who modelled its spectral energy distribution (SED).  \citet{1995A&AT....6..251M} derived two best models, a single B0V star with a 1150~K dust shell and a binary system with a B0V + K0II pair with a 1050~K dust shell.  \citet{1999ApJ...527..345B} fitted a Kurucz model plus thin dust model to their pre-outburst data and found the SED to be consistent with a $\geq 22000$~K B2 or earlier star plus a large mass of hot dust. The best-fitting model by \citet{2000A&A...356...50C} included a 30000~K  Kurucz model plus 1500K hot dust. \citet{2002ApJ...565.1169R} estimated the radius of the inner edge of the dust region to be 13~AU$<r_{shell}<$52~AU, assuming a dust temperature T$_{dust}$=1350~K and a spherical distribution of the material. With a distance to CI Cam of 5~kpc, this corresponds to an angular radius in the 2.6 to 10.4~mas range. The first interferometric observations of CI Cam's dust region were obtained with the Infrared Optical Interferometer Array (IOTA) shortly after the 1998 outburst \citep{2003SPIE.4838..202M}. Their best-fitting uniform disc models gave diameters of $5.6\pm 0.4$ mas and $7.4 \pm 0.4$ mas in the H and K' spectral bands respectively.

In this paper we present the first resolved parametric imaging model of the hot dust region of CI Cam obtained using the high angular resolution provided by the IOTA and the PTI interferometers. We introduce our interferometry and photometry observations in section~\ref{sec:Observations}. We investigate the properties of the circumstellar environment of CI Cam through the modelling of its spectral energy distribution (SED) based on photometry/spectroscopy data contemporary with our interferometry data in section~\ref{sec:The spectral energy distribution}.  We discuss our interferometry data modelling in section~\ref{sec:The interferometry data}. And finally draw an updated picture of the circumstellar envelope of CI Cam and discuss future work in section~\ref{sec:Summary and future work}.

\section{Observations}
 \label{sec:Observations}
\begin{figure}
\begin{center}
\includegraphics[width=80mm]{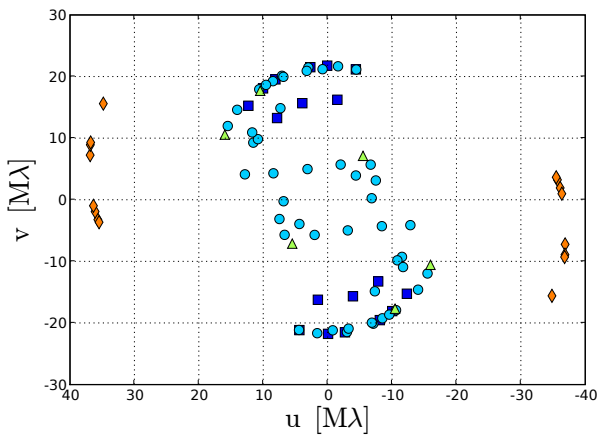} 
\caption{The (u,v)-plane coverage of CI Cam. Squares: IOTA 2T 1998 ; Circles: IOTA 3T 2004, Triangles: IOTA 3T 2005; Diamonds: PTI 2006.}
\label{fig:uv}
\end{center} 
\end{figure}

In this paper we combine near-infrared interferometric data acquired with the free-space beam combiner \citep{1999PhDT........25M,2001ApJ...546..358M} and IONIC-3 beam combiner \citep{2003SPIE.4838.1099B} on the IOTA interferometer \citep{1998SPIE.3350..848T,2003SPIE.4838...45T} with data obtained with the Palomar Testbed Interferometer \citep{1999ApJ...510..505C}. The observations cover the period from September 1998 to December 2006. Figure~\ref{fig:uv} shows the (u, v)-coverage obtained with IOTA and PTI for the science target. The stars we used as calibrators are given in Table~\ref{tab:logIOTA3T.PTI}.

\subsection{IOTA2T free-space combiner \label{sec:IOTA2T free-space combiner}}
These observations were carried out at IOTA using the two telescopes available at the time, located on the North-East arm. The measurements were taken in the near-infrared $ H (\lambda_{0} = 1.65~\mu m, \Delta \lambda = 0.30~\mu m)$ and $ K' (\lambda_{0} = 2.16~\mu m, \Delta \lambda = 0.32~\mu m)$ spectral bands. A description of the observations can be found in  \citet{2003SPIE.4838..202M} and \citet{1998AAS...193.5206T}. The telescope configuration provided a projected baseline length in the 32-36~m range. The starlight beam from each telescope were combined at a 50-50 beam splitter. The output of the beam splitter were two combined beams which were focused on two pixels of a NICMOS3 detector. The interference fringes were produced by modulating the optical path difference (OPD) between the two beams. The beam combiner and NICMOS detector are described in \citet{2005ApJ...620..961M, 2001ApJ...546..358M, 1999PASP..111..238M}. 

\subsection{IOTA3T IONIC-3}
Our second set of interferometric data was obtained using the upgraded three-telescope IOTA interferometer \citep{2003SPIE.4838...45T} with the new integrated optics beam combiner IONIC3 \citep{2003SPIE.4838.1099B}. We used three different baseline configurations that were obtained by moving the A and C telescopes along the North-East arm and the B telescope along the South-East arm.  CI Cam was observed using a standard H band filter.

The three beams are fed to the IONIC3 beam combiner which provides a pairwise combination by means of integrated optics (IO). The IONIC3 combiner contains single-mode waveguides. The three inputs are split with three "Y" junctions to provide a pairwise beam combination with another set of three couplers. A detailed description of the IO component is found in \citet{2003SPIE.4838.1099B}. The six light beams emerging from the combiner are focused onto 6 separated pixels of a PICNIC Rockwell detector \citep{2004PASP..116..377P}. The optical paths are equalised by an automated fringe packet tracker  described in \citet{2005ApOpt..44.5173P}.

Target observations are interleaved with identical observations of calibrator stars that are unresolved for the interferometer. The observing procedure and data reduction are described in \citet{2004ApJ...602L..57M}. A log of the observation is given in table \ref{tab:logIOTA3T.PTI}.

\begin{table*}
\centering
 \begin{minipage}{130mm}
\caption{Observation log for CI Cam. The IOTA telescope configuration refers to the location of the A, B, C telescopes along the NE, SE and NE arms.}
\label{tab:logIOTA3T.PTI}
\begin{tabular}{ll|cc|ll}
\hline
\hline
\multicolumn{2}{c|}{Date} & & & \multicolumn{2}{|c}{Calibrator Names}\\
\multicolumn{2}{c|}{(UT)} & Instrument & Telescope configuration & \multicolumn{2}{|c}{(Adopted UD diameter)}\\
\hline
1998 Sep 27 & & IOTA 2T & A35-B15 & HR1313 (1.6 $\pm$ 0.8 mas) $^{a}$& \\
1998 Sep 28 & & IOTA 2T & A35-B15 & HR1255 (1.2 $\pm$ 0.7 mas)$^{a}$& \\
1998 Nov 04 & & IOTA 2T & A35-B15 & HR1255 (1.2 $\pm$ 0.7 mas) $^{a}$& \\
2004 Nov 26-27 & & IOTA 3T & A35-B15-C00 & HD27322 (0.24 $\pm$ 0.10 mas)  $^{a}$& \\
2004 Dec 02  & & & & HD25948 (0.30 $\pm$ 0.10 mas) $^{a}$&\\
 & & & & HD26553 (0.35 $\pm$ 0.10 mas)  $^{a}$&\\
 & & & & HD27224 (0.87 $\pm$ 0.01 mas) $^{b}$&\\
2004 Dec 12-14 & & IOTA 3T & A28-B10-C00 & HD27224; HD26553 &\\
2004 Dec 16 & & IOTA 3T & A28-B05-C10 & HD27224; HD26553 &\\ 
2005 Nov 25 & & IOTA 3T & A35-B15-C00 & HD27224; HD26553 &\\
2006 Oct 31 & & PTI & N-W & HD26764 (0.5 $\pm$ 0.7 mas) $^{a}$& \\
 & & & & HD25948; HD 26553 &\\
2006 Nov 11 & & PTI & N-W & HD26553; HD25948; HD 26764 &\\
2006 Nov 13 & & PTI & N-W & HD26553; HD25948 &\\
2006 Nov 25 & & PTI & N-W & HD26553; HD25948; HD 26764 &\\
2006 Dec 12 & & PTI & N-W & HD26553; HD25948; HD 26764 &\\ 
\hline
\end{tabular} 
\newline $^{a}$ UD diameters calculated with the fBol module of getCal (Michelson Science Center, California Institute of Technology, http://msc.caltech.edu
\newline $^{b}$ UD diameter from \citet{2005A&A...433.1155M}
\end{minipage}
\end{table*}

\subsection{PTI}
Our third set of data was obtained using the Palomar Testbed Interferometer, a three-element long-baseline interferometer that combines two of its three apertures to make measurements of fringe visibility for astronomical sources. For this particular observation, the measurements were taken with five narrow bands across the K band that were centred at 2.009, 2.106, 2.203, 2.299, and 2.396~$\mu$m. For all of these observations, PTI's 86 m north-west baseline was used; details on PTI can be found in Colavita et al. (1999).

CI Cam was observed, along with the unresolved calibration sources HD 26553, HD 25948, and HD 26764, on 5 nights. A log of the observation is given in Table \ref{tab:logIOTA3T.PTI}. The PTI composite spectrometer data were calibrated using the WebCalib\footnote{http://mscweb.ipac.caltech.edu/webCalib/} web interface provided by the Michelson Science Center.

\subsection{Photometry data}
\label{Photometric_data}
The photometric observations used to construct CI Cam SED model were taken from the literature and from observations carried out at the Michigan-Dartmouth-MIT (MDM) Observatory in 2004 and 2005. The data span the wavelength range from the UV to the infrared: \citet{2002yCat..80790309B}, \citet{2000A&A...356...50C},  Two Micron All Sky Survey (2MASS) \citep{2006AJ....131.1163S}, Infrared Astronomical Satellite (IRAS) \citep{1988iras....7.....H} and Spitzer (IRS spectra program ID 20702). When computing the average flux for each wavelength, we excluded measurements taken during the outburst and used post-outburst data when available.\\
The MDM photometry measurements were taken with the 2.4~m Hiltner telescope at the MDM observatory in the UBVRIJHK spectral bands in Nov-Dec 2004 and Dec 2005. U,B,V,R,I data were taken with the MDM 8K\footnote{http://www.astro.columbia.edu/~arlin/MDM8K/index.html} camera equipped with Johnson U,B.V,R, Kron-Cousins I filters. J, H and K observations were done with the TIFKAM\footnote{http://www.astronomy.ohio-state.edu/mosaic/index.html} infrared camera. Table~\ref{tab:MDM1} summarises our measurements.

\begin{table*}
\caption{UBVRIJHK photometry of CI Cam obtained by A. Tannirkulam at the MDM observatory.}
\begin{center}

\label{tab:MDM1}
\begin{tabular}{llcccccccc}
\hline
\hline
\multicolumn{2}{c}{Date} & U & B & V & R & I & J & H & K\\
\multicolumn{2}{c}{(UT)} & & & & & & & \\
\hline
2004 Nov 27 & & 12.13$\pm$0.07 & 12.41$\pm$0.06 & 11.77$\pm$0.05 & 10.79$\pm$0.05 & \\
2004 Dec 1 & & & & & & 9.99$\pm$0.04 & 7.20$\pm$0.10 & 5.68$\pm$0.10 & 4.44$\pm$0.10\\
2005 Dec 17& & & & & & & 7.01$\pm$0.05 & 5.64$\pm$0.05 & 4.35$\pm$0.05\\
\hline
\end{tabular} 
\end{center}
\end{table*} 

\section{Modelling the dust shell of CI Cam}

\subsection{Comparison of the observed SED and visibilities with spherical dust shell models \label{sec:The spectral energy distribution}}
We used the radiative transfer code DUSTY developed by \citet{1997MNRAS.287..799I} in order to compute the SED for our models. For the sake of comparison, we constructed SED models for both the pre-outburst and post-outburst data. 

We computed the SEDs with the stellar radiation being described by a black body of temperature T$_{eff}$. The central star is surrounded by a spherical dust shell assumed to begin at R$_{in}$, inner boundary of the shell and extend up to R$_{out}$. The temperature at the inner boundary of the shell is T$_{dust}$. The dust grain type is set to amorphous carbon since no silicate features appear to be present neither in the IRAS Low Resolution Spectra (LRS) nor in the Spitzer Infrared Spectrograph (IRS) spectrum (D. S. Moon 2007, private communication). The dust grain size distribution is a power law with minimum and maximum size boundaries at a$_{min}$ and a$_{max}$. The dust shell density profile follows a single power-law with $\rho \propto r^{-p}$. The output parameters are: the angular diameter of the star $\theta_{star}$, the angular inner radius of the dust shell R$_{in}$, the percentage of stellar flux F$_{star}$ in the H and K spectral bands, the interstellar extinction E$_{B-V}$, the stellar luminosity L$_{star}$ and the dust opacity in the H and K spectral bands $\tau _{H}$ and $\tau _{K}$. A rough estimate of the dust mass M$_{dust}$ is calculated using the equations which relates $\tau _{V}$ and  M$_{dust}$ derived by \citet{2005A&A...434..849H}. The values given in table~\ref{tab:sed_params} should be used with caution as the dust mass uncertainties are high due to the inaccurate distance and complex dust geometry. See \citet{2005A&A...434..849H} for a discussion on the dust mass uncertainties. The input parameters and best-fitting model parameters for both pre- and post-outburst data are summarised in table~\ref{tab:sed_params}. 

\begin{figure*}
\begin{center}
\begin{tabular}{cc}
\includegraphics[height = 6cm]{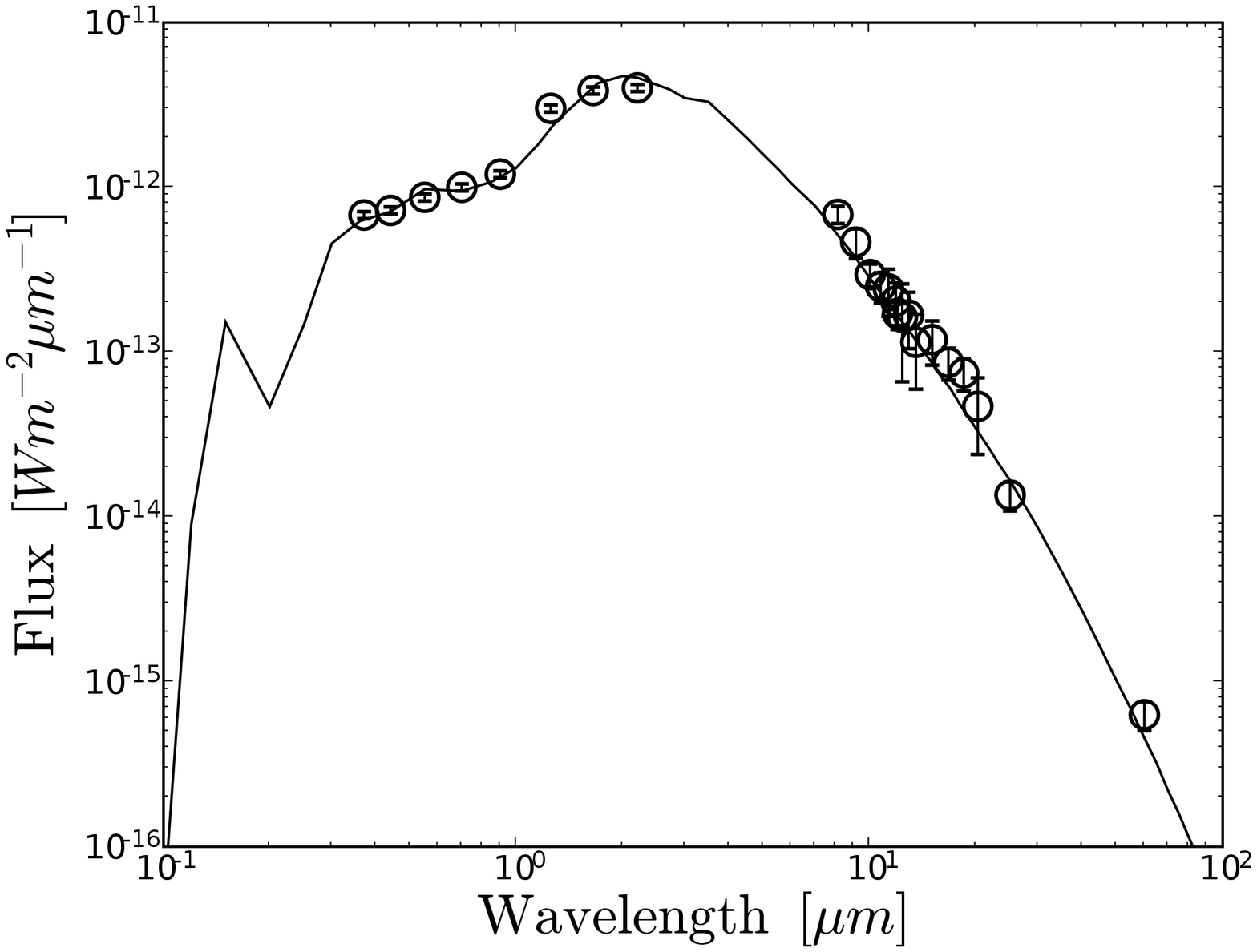}  & \includegraphics[height = 6cm]{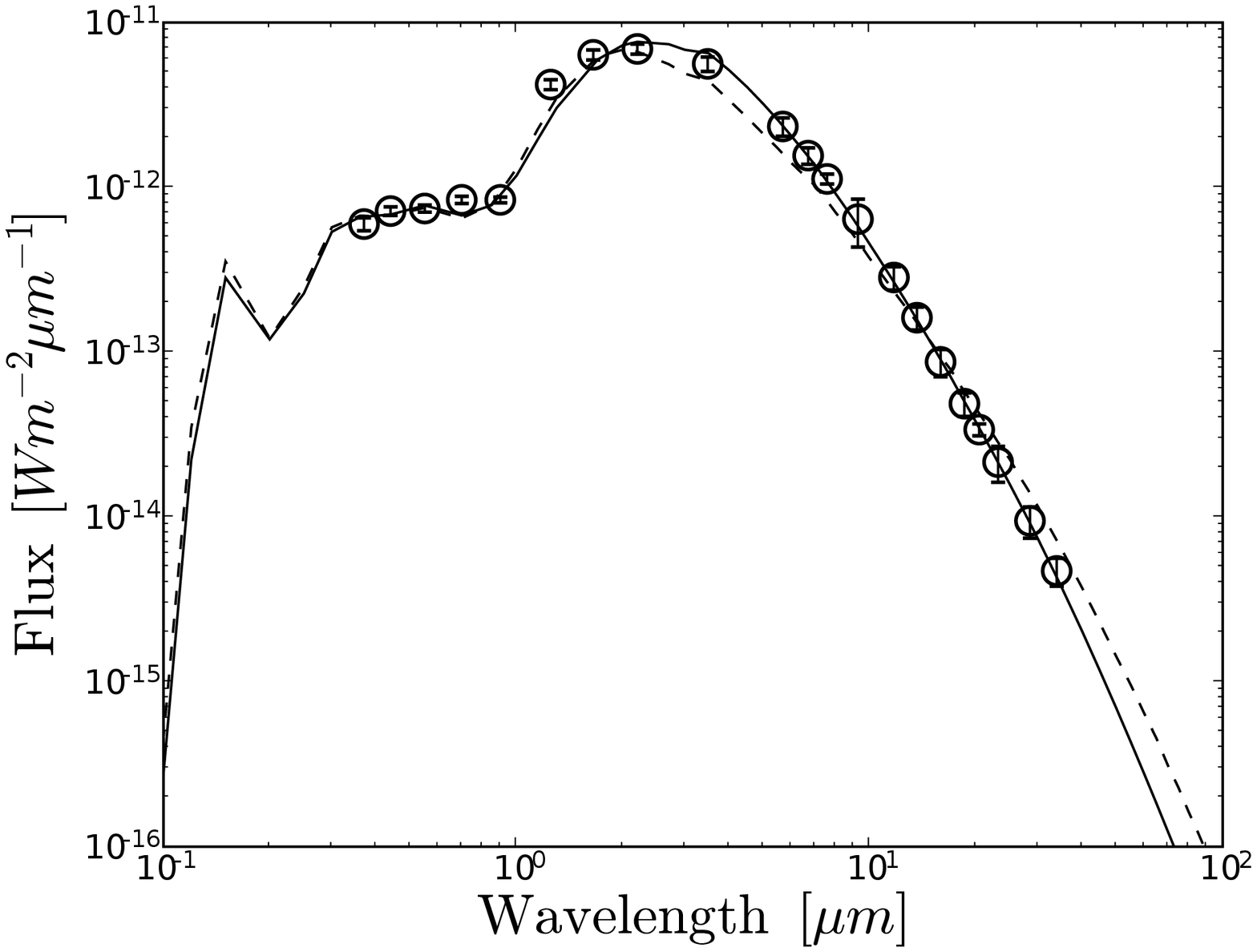}\\
\end{tabular} 
\caption{ \label{fig:sed_fit} Spectral energy distribution data and DUSTY model of CI Cam. Left hand panel: pre-outburst data; the model uses the parameters T$_{eff}$, T$_{dust}$ and $p$ from \citet{2000A&A...356...50C}. Right hand panel: post-outburst data; the dashed line model uses the parameters from \citet{2000A&A...356...50C}; the plain line uses our new set of parameters.}
\end{center}
\end{figure*}

\begin{figure*}
\begin{center}
\begin{tabular}{cc}
\includegraphics[height = 6cm]{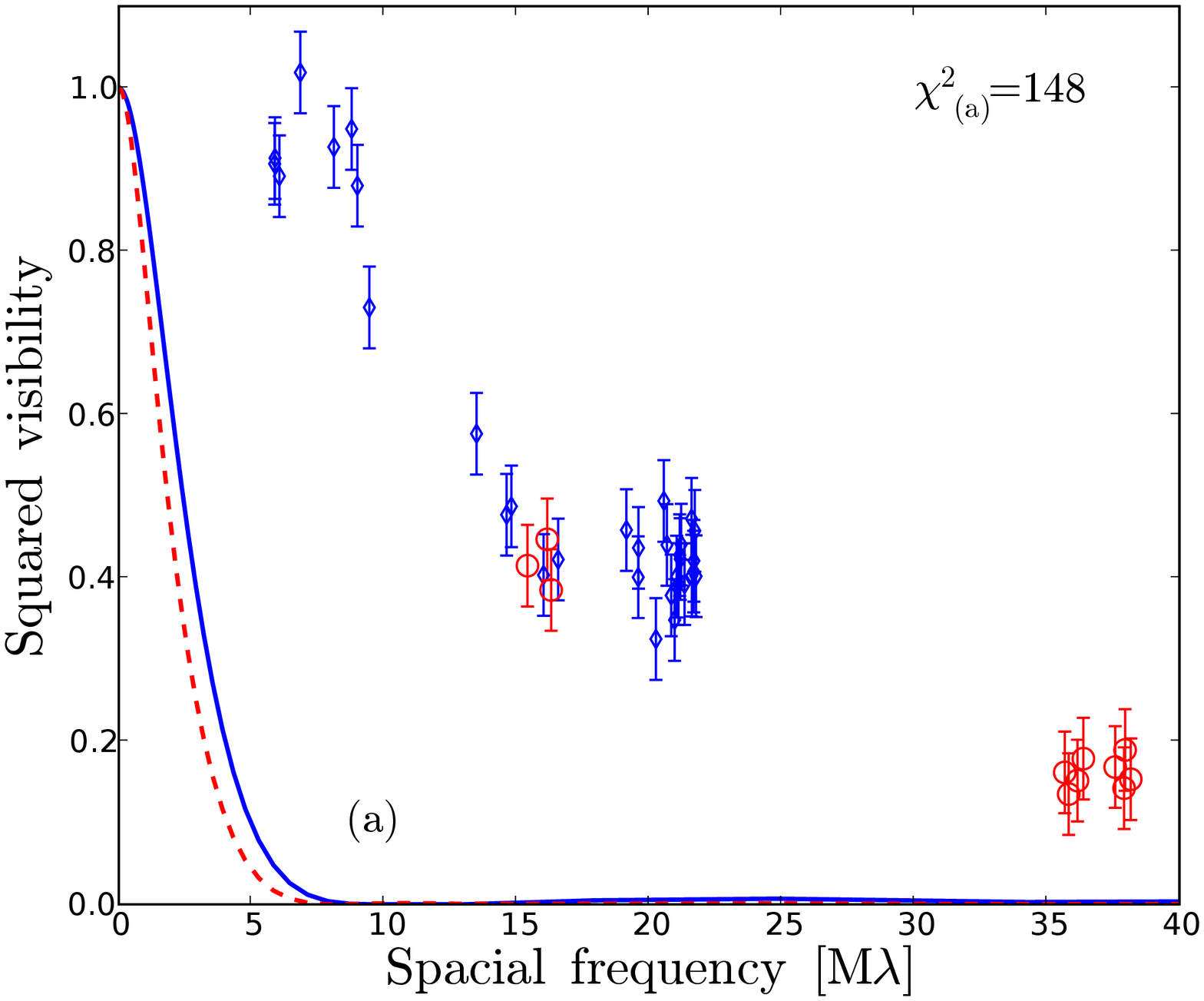}&\includegraphics[height = 6cm]{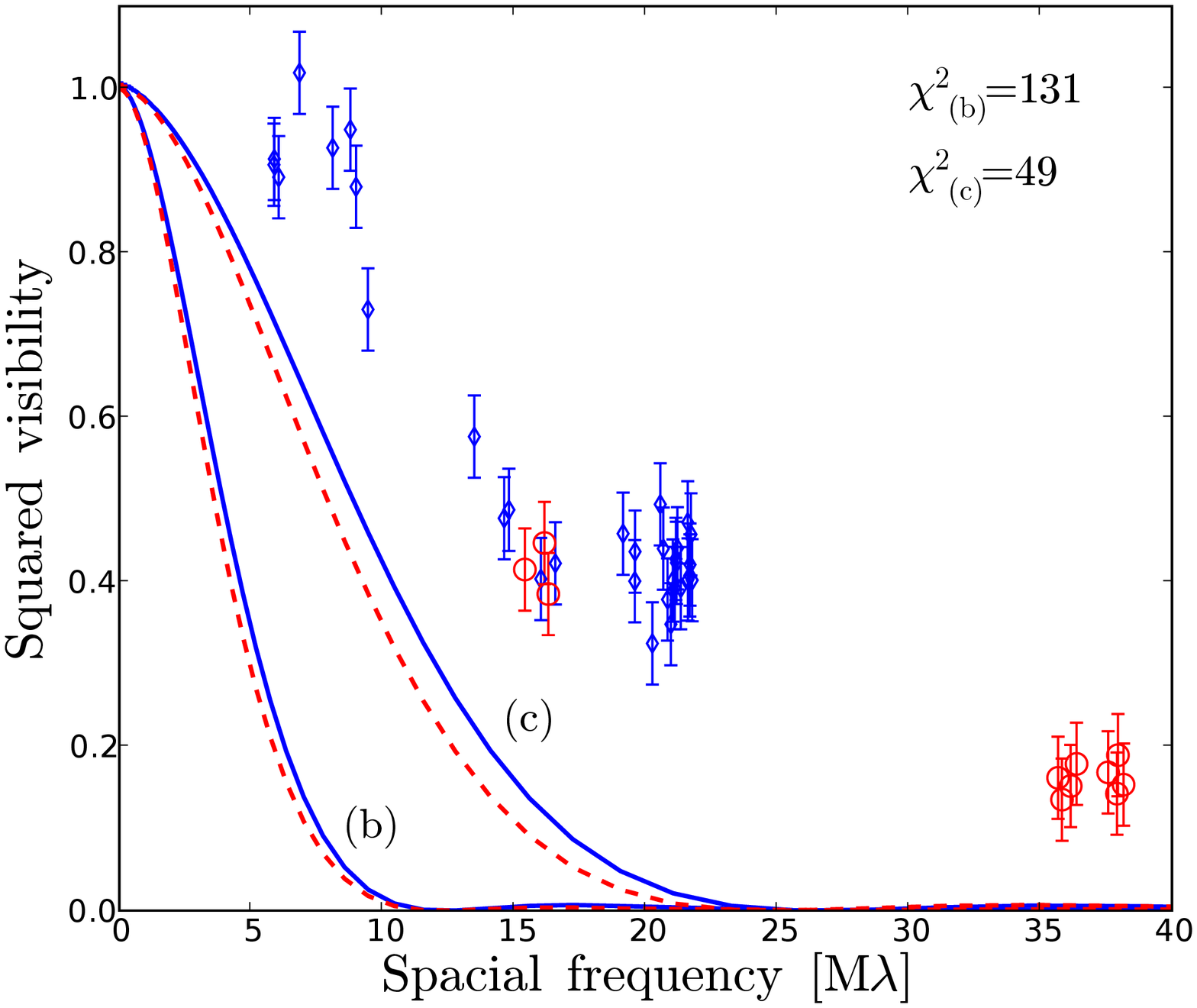}\\
(\ref{fig:vis_sed}a) Pre-outburst, Spherical dust shell & (\ref{fig:vis_sed}b) Post-outburst, Spherical dust shell\\
\end{tabular} 
\caption{\label{fig:vis_sed}The visibility curves for the DUSTY models are plotted alongside the IOTA/PTI uv-averaged visibility data points. The diamonds (plain lines) represent the H band data (models) while the open circles (dashed lines) represent the K band data (models). The left hand panel corresponds to the pre-outburst data, SED model (a) while the right hand panel shows the model visibilities for the post-outburst data for both model (b) and (c).}
\end{center}
\end{figure*}

Our pre-outburst model (a) is constructed from averaged photometry data presented in \citet{1995A&AS..112..221B},  [12, 25, 60]~$\mu m$ IRAS fluxes \citep{1988iras....7.....H} and IRAS LRS spectrum\footnote{http://www.iras.ucalgary.ca/$\sim$volk/getlrs\_plot.html}. As mentioned by \citet{2000A&A...356...50C} the IRAS spectrum is quite noisy, and was averaged before being combined with the photometry data set. Model (a) uses \citet{2000A&A...356...50C} best-fitting parameters as input to DUSTY. Their SED models, also computed with DUSTY, include a T$_{eff}$=30000~K  Kurucz model plus T$_{dust}$=1500~K hot carbon dust with a standard dust grain size distribution and density power law index $p$=1.75 for both the pre-outburst and post-outburst SEDs. Both models differ only in the value of the overall optical depth of the dust envelope. The best-fitting SED model (a) is shown with the pre-outburst photometry data in figure~\ref{fig:sed_fit}. The model fits the outburst data well, however, the computed visibilities fail to reproduce the interferometric observations as shown in figure~\ref{fig:vis_sed}a. The much too small visibilities implies either that the size of the dust shell, R$_{in}$ = 9.7~mas, as determined with the SED model fitting is larger than the size measured with interferometry or that the dust shell was larger prior to the outburst. Since no pre-outburst interferometry data are available we can not rule out the latter.

We applied the same procedure to post-outburst photometry data. Our post-outburst models are constructed from averaged MDM (see table~\ref{tab:MDM1}) and 2mass \citep{2006AJ....131.1163S} photometry data as well as a Spitzer IRS spectrum. The Spitzer spectrum was downloaded from the Public Post-BCD Data Webserver, IRSA\footnote{http://irsa.ipac.caltech.edu/applications/Spitzer/Spitzer/}.  The Spitzer spectrum was binned in order to obtain an homogeneous distribution of data points across our entire wavelength range. Our first approach was to apply the input parameters from model (a). Our best-fitting model (b) is shown in figure~\ref{fig:sed_fit}. The model fails to reproduce the photometry data accurately and the size of the dust shell found with DUSTY, R$_{in}$ = 6.6~mas, is too large for the model visibilities to match our measurements as seen in figure~\ref{fig:vis_sed}b. Our second approach consisted in running DUSTY for a range of T$_{eff}$ = [15000 - 30000]~K, a range of T$_{dust}$ = [1000-1600]~K, a power law parameter $p$ between 1.5 and 2.5 and various grain size distributions. Our motivation was to derive a model with a dust inner radius consistent with the size measured with the IOTA interferometer. A lower effective temperature and larger dust grain size would be expected to generate a dust shell with a smaller inner radius. Our best-fitting model (c) comprises of a T$_{eff}$ = 24000~K star and a T$_{dust}$ = 1550~K dust shell. Our grain size distribution includes much larger grains up to a size of a$_{max}$ = 5.75$~\mu m$ with a much steeper density power low distribution where $p$ = 2.35. The SED curve for model (c) is shown in figure~\ref{fig:sed_fit}. The flux originating from the star accounts for only 1.8\% of the total flux at $\lambda=1.65\;\mu$m (0.6\% of the total flux at $\lambda=2.2\;\mu$m), the main contribution to the total flux being the emission from the hot dust. The SED model is in very good agreement with the photometry data with the exception of the J band data point where the modelled flux is lower than its actual value. The model visibility curve still fails to reproduce the interferometry data with model (c) dust shell size R$_{in}$ = 3.3~mas (see figure~\ref{fig:vis_sed}b). The main reason for the remaining discrepancy is that a spherical dust environment fails to accurately describe the geometry of the dust shell surrounding CI Cam as is shown in section~\ref{sec:The interferometry data}.

\begin{table}
\caption{\label{tab:sed_params} Input and output parameters of the SED model fitting. Sizes in AU and dust masses were derived assuming that CI Cam is at a distance d=5kpc.}
\centering 
 \begin{tabular}{lccc}
\hline
\hline
& Pre-outburst & \multicolumn{2}{c}{Post-outburst} \\
& Model (a) & Model (b) & Model (c)\\
\hline
\multicolumn{4}{c}{Input parameters}\\
\hline
T$_{eff}$[K] & 30000 & 30000 & 24000 \\
T$_{dust}$[K] & 1500 & 1500 & 1550 \\
$p$ &1.75 & 1.75 & 2.35\\
a$_{min}$-a$_{max}$ [$\mu m$] & 0.005-0.25 & 0.005-0.25 & 0.005-5.75\\
\hline
\multicolumn{4}{c}{Output parameters}\\
\hline
$\theta_{star}$ [mas] & 0.026 & 0.020 & 0.030\\
D$_{star}$ [AU] & 0.15 & 0.10 & 0.13\\
R$_{in}$ [mas] & 9.7 & 6.6 & 3.3 \\
R$_{in}$ [AU] & 48 & 33 & 16 \\
F$_{star}$ [\%], 1.65~$\mu$m & 5.5 & 1.8 & 1.8 \\
F$_{star}$ [\%], 2.2~$\mu$m  & 2.3 & 0.7 & 0.6\\
E$_{B-V}$ & 1.16 & 0.92 & 0.87 \\
L$_{star}$ [L$_{\sun}$] & 10$^{5.3}$ & 10$^{4.9}$ & 10$^{4.8}$ \\
$\tau_{H}$ & 0.02 & 0.07 &0.38 \\
$\tau_{K}$ & 0.01 & 0.04 & 0.32\\
$\tau_{V}$ & 0.07 & 0.23 & 0.63\\
M$_{dust}$ [M$_{\sun}$] & 3$\times$10$^{-6}$ & 5$\times$10$^{-6}$ & 8$\times$10$^{-7}$\\
\hline
\end{tabular}

\end{table} 

\subsection{Skewed ring model derived from the interferometry data \label{sec:The interferometry data} }
\begin{figure*}
\begin{center}
\begin{tabular}{cc}
\includegraphics[height = 6cm]{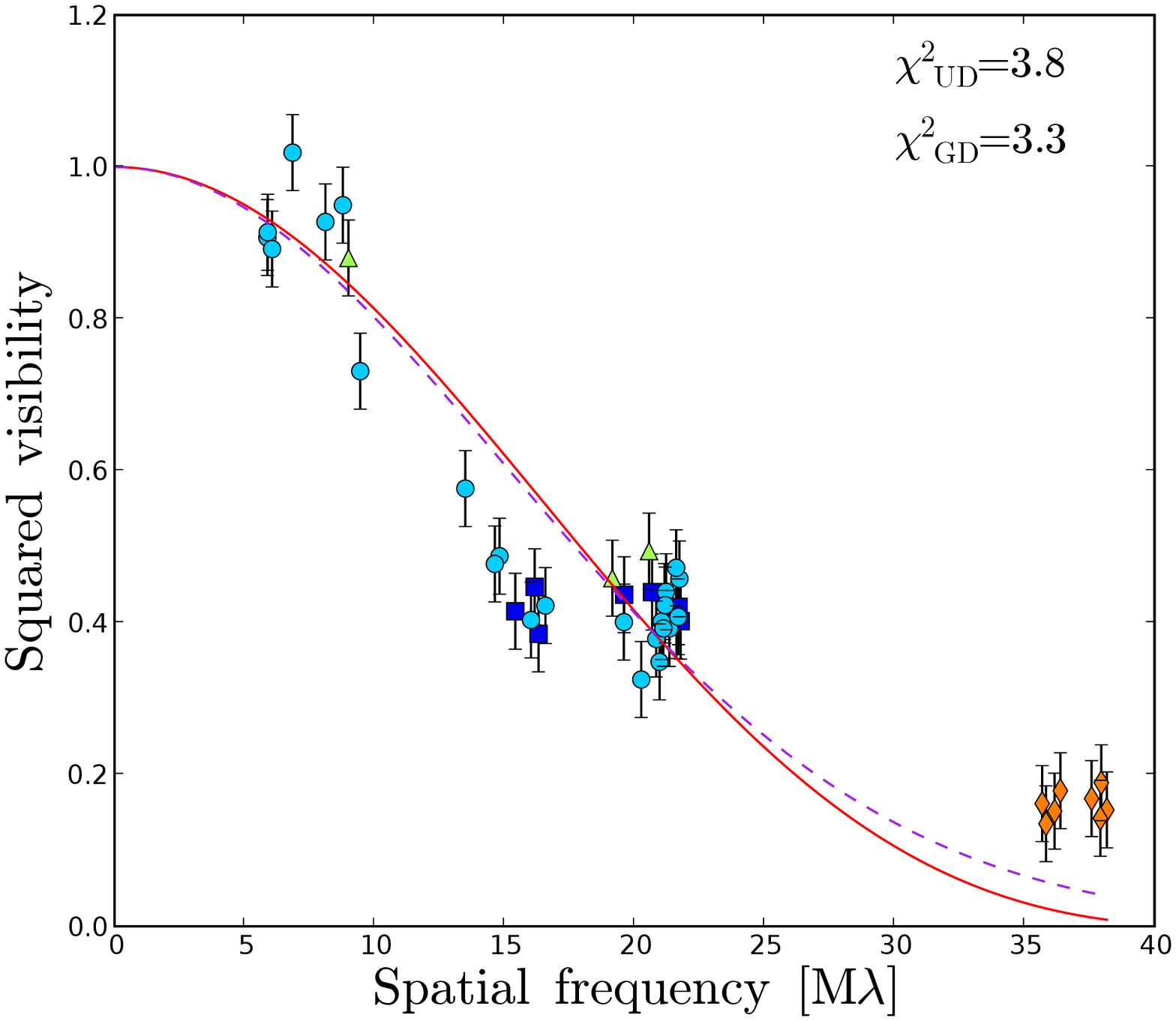} & \includegraphics[height = 6cm]{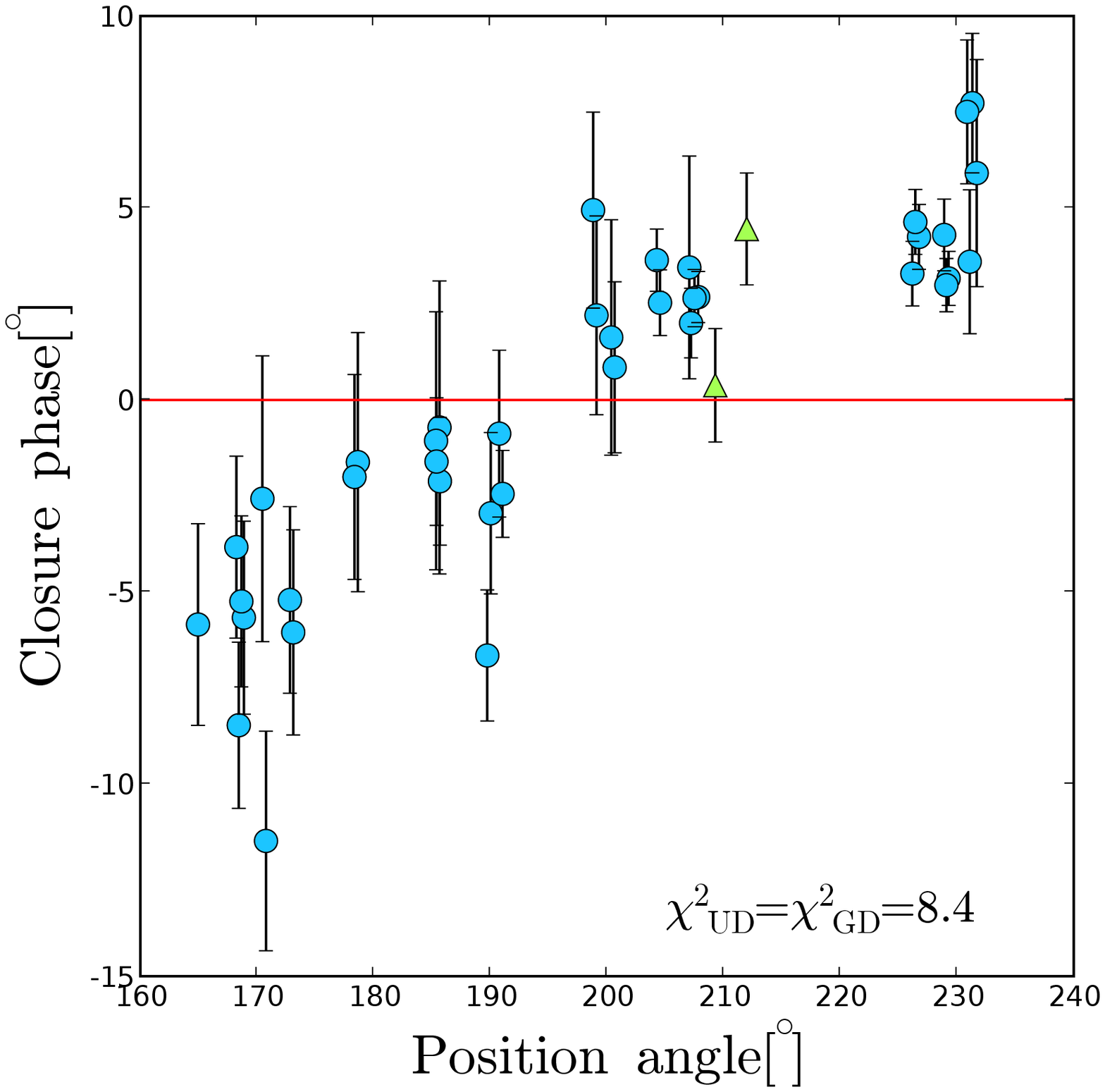} \\
(\ref{fig:vis}a) Visibility data with simple Gaussian and uniform disc models & (\ref{fig:vis}b) Closure phase data \\
\end{tabular} 
\caption{\label{fig:vis} Left panel: uv-averaged visibility data points measured with the IOTA and PTI interferometers. The solid line and dashed line represent the uniform disc ($\theta_{UD} = 5.93\pm 0.09$ mas)  and Gaussian disc ($FWHM = 3.3 \pm 0.05$ mas) best-fitting models respectively. Right panel: closure phase data measured with IOTA3T. Plotting symbols identify the separate data sets. Squares: IOTA 2T 1998 ; Circles: IOTA 3T 2004, Triangles: IOTA 3T 2005; Diamonds: PTI 2006 }
\end{center}
\end{figure*}

The interferometric  observations were carried out in the H and K spectral bands. These wavelengths are well suited to measure the size and study the geometry of the hot dust surrounding CI Cam. IOTA visibility data taken in 1998, 2004 and 2005, as shown in figure~\ref{fig:vis}a, agree well with each other which means that the structure of the dusty environment of the star, as seen by the interferometer, has not changed over the years. This allows us to combine our data sets and use them as one in our model fitting. From the shape of the visibility curve shown in figure~\ref{fig:vis}a, a uniform or Gaussian disc models are clearly incompatible with the observations. We measured non-zero closure phase (see figure~\ref{fig:vis}b) which indicates that the source brightness distribution in the H band is asymmetric. No closure phases measurements were taken in the K band. Although we display uv-averaged visibility data in the plots throughout this paper, we used the original data
sets in our computations. 

\begin{figure*}
\begin{center}
\begin{tabular}{cc}
\includegraphics[height = 6cm]{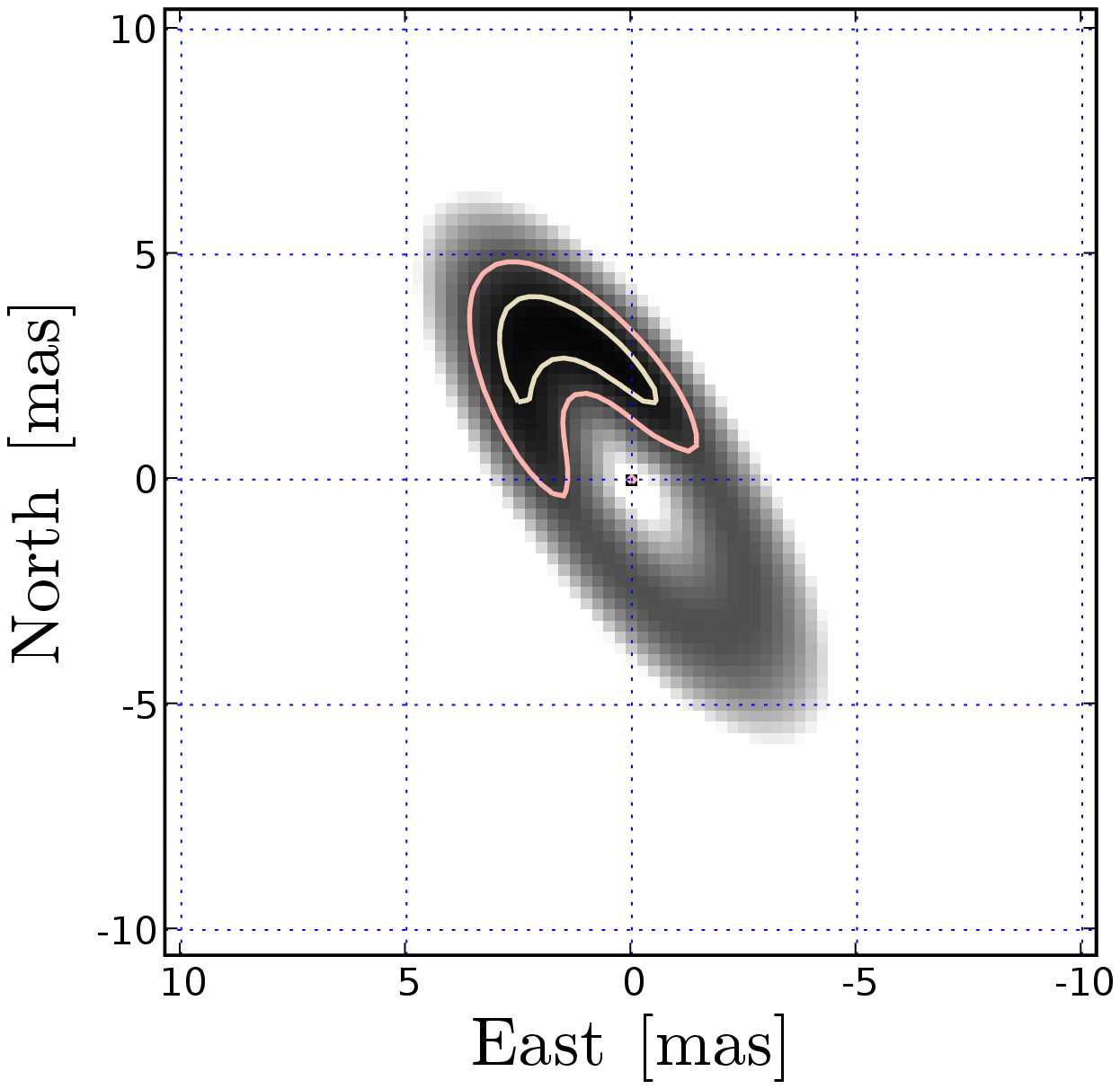} & \includegraphics[height = 6cm]{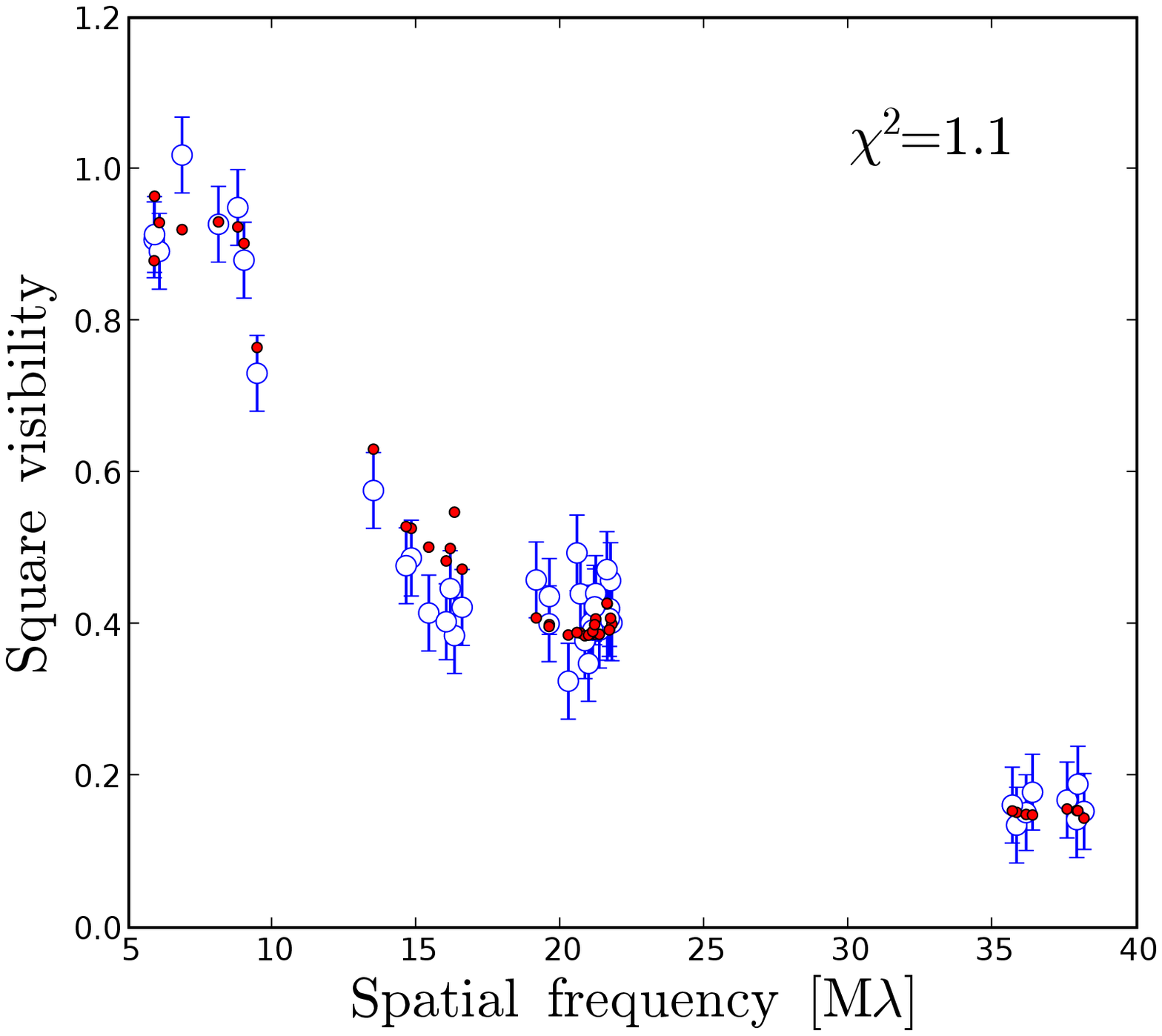} \\
\includegraphics[height = 6cm]{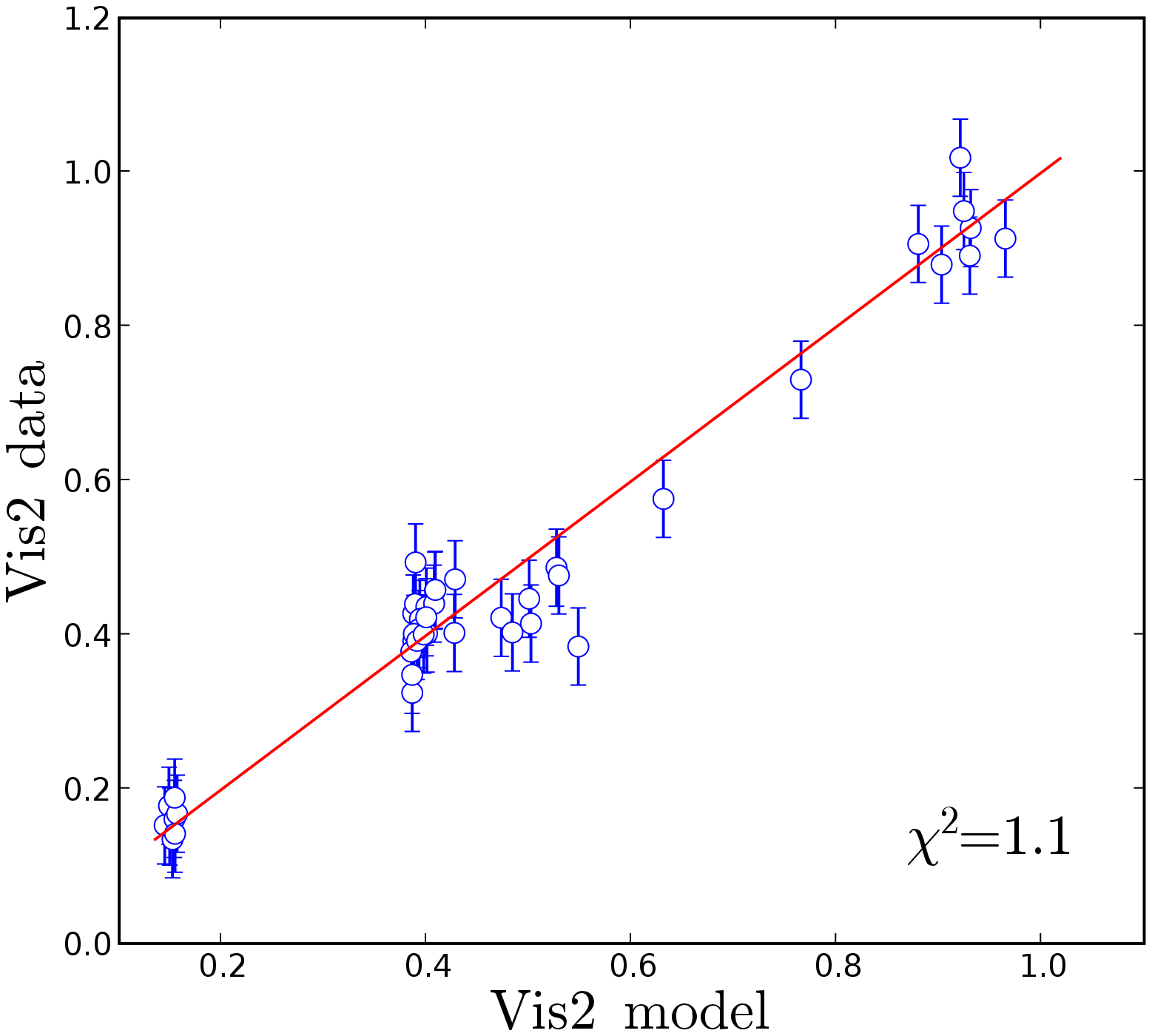} & \includegraphics[height = 6cm]{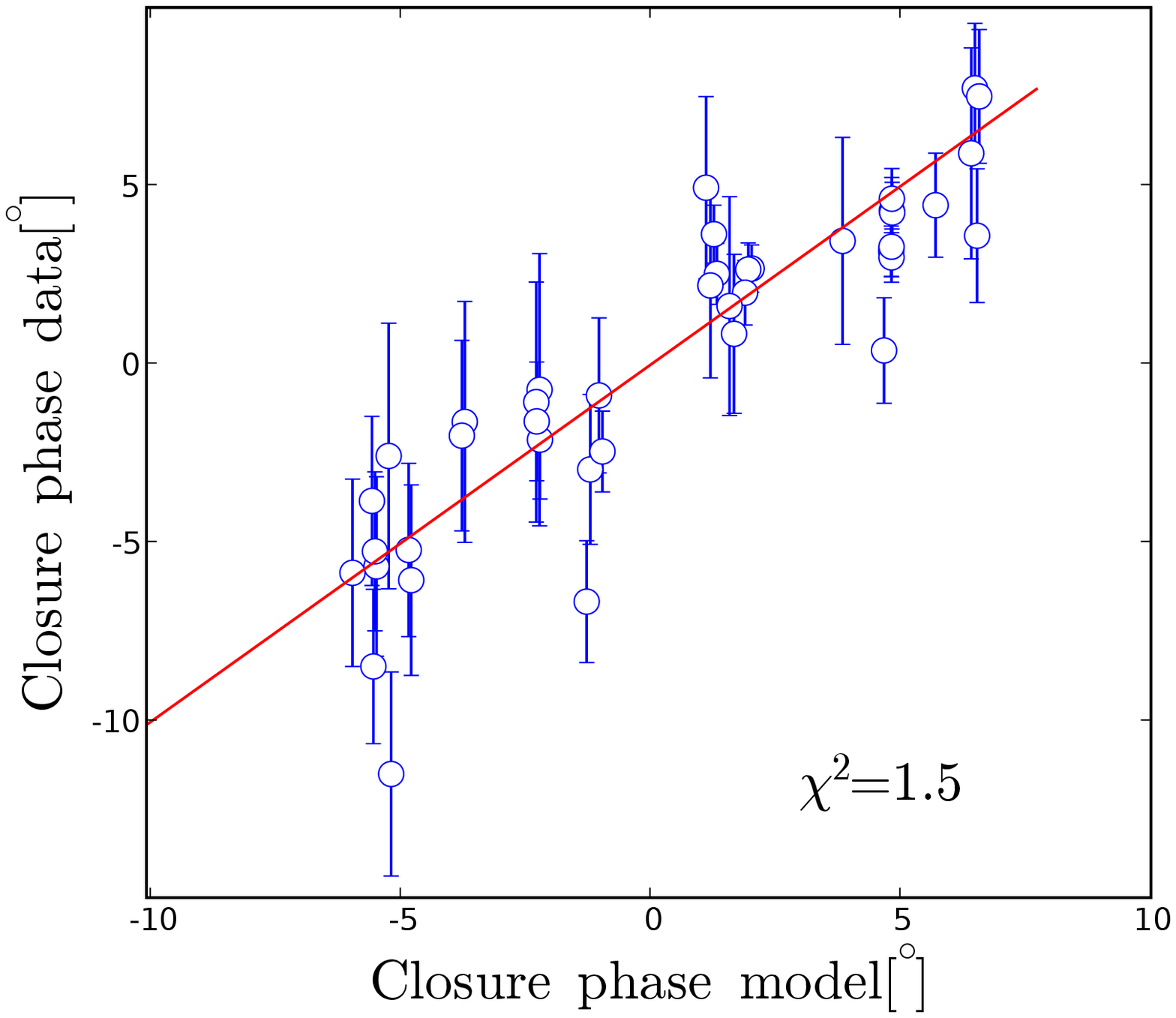} \\
\end{tabular} 
\caption{\label{fig:best_fit_model} Top left panel shows the best-fitting Gaussian skewed ring model to the interferometry data. The contours in the image correspond to 0.1, 0.5, 0.9 of the peak intensity. Top right panels shows the model square uv-averaged visibilities (plain circles) along the interferometry data (open circles). Bottom left panel compares the observed uv-averaged visibility data with the model data. Bottom right panel compares the observed closure phases with the model closure phases.}
\end{center}
\end{figure*}

We tested several geometric models against the data. Our best-fitting model is a Gaussian skewed ring, where the ring emission is modulated as a function of the azimuth, plus a point source that accounts for the unresolved contribution of the central star. Skewed disc models were previously discussed by \citet{2006ApJ...647..444M} in the context of the modelling of  the near-infrared emission from the dust discs of young stellar objects. This class of models assumes that dust discs seen in the near-infrared will appear as a ring. The azimuthal intensity modulation would either account for the inner edge being seen at an angle or for inhomogeneities in the dust distribution. If seen at an angle the ring would be elliptical, not circular. The parameters of the model are: the flux of the ring, the flux of the point source, the ring major axis, the ring axis ratio, the position angle of the ring, the skewness, and the skewed angle which is the position angle of the area of the ring with the greatest skewed brightness. All angles are given east of north. Our model uses the averaged H and K stellar flux derived from the SED fit to constrain the flux of the two components. The best-fitting model is presented in figure~\ref{fig:best_fit_model} and the resulting parameters are listed in table~\ref{tab:mpars}. 

We find that CI Cam hot dust can be modelled as a skewed elliptical Gaussian ring where the semi-major axis matches the inner radius of the dust found for our DUSTY model (c). This is quite significant since it means that the location of the hot dust as measured by interferometry is the same as the location of the hot dust as predicted with DUSTY for this specific model (c) which is characterised by a lower T$_{eff}$, a larger dust grain population and a somewhat lower luminosity. Assuming the distance to CI Cam is 5~kpc, we find the ring major axis to be  37.9$\pm$1.2~AU. The ellipticity of the hot dust region argues in favour of a disc-shaped dust distributions rather than a spherical dust distribution. The skew angle is different from the position angle of the ring itself by about 18$^{\circ}$ which shows that the skewness doesn't come from an inner rim seen at an angle, where the difference in angle would be 90$^{\circ}$, but rather arises from an asymmetric distribution of the hot dust around CI Cam. The asymmetry could arise from the formation of clumps during the condensation of the dust grains or, alternatively the asymmetric geometry of the hot dust could be generated by the presence of a hot undetected companion that would evaporate some of the dust in a non homogeneous way in the inner dust region of CI Cam. We also find that the orientation of the dust shell of CI Cam is similar to the orientation of the radio images \citep{2004ApJ...615..432M} where they find the position angle of the radio emission to be about 20$^{\circ}$ east of north. 

\begin{table}
\begin{center}
\caption{Parameters of CI Cam Gaussian skewed ring model.}
\begin{tabular}{lr@{$\pm$}l}
\hline
\hline
& \multicolumn{2}{c}{Model parameters} \\
\hline
Ring flux [fraction total flux] & \multicolumn{2}{c}{0.988} \\
Point source flux [fraction total flux] & \multicolumn{2}{c}{0.012}  \\
Ring major axis [mas] & 7.59 & 0.29  \\
Ring major axis [AU], d=5~kpc & 37.9 & 1.2  \\
Axis ratio & 0.39 & 0.03  \\
Ring position angle [$^{\circ}$] & 35 & 2  \\
Skewness & 0.86 & 0.01  \\
Skewed angle [$^{\circ}$] & 17 & 3  \\
$\chi ^{2}_{vis^{2}}$ &  \multicolumn{2}{c}{1.1}  \\
$\chi ^{2}_{cp}$ &  \multicolumn{2}{c}{1.5}  \\
$\chi ^{2}_{vis^{2},cp}$ &  \multicolumn{2}{c}{1.1}  \\
\hline
\end{tabular} 
\label{tab:mpars}
\end{center}
\end{table} 

\section{Summary and future work}
\label{sec:Summary and future work}
Our SED modelling doesn't yield to any conclusion regarding the spectral classification of CI Cam. According to \citet{1996imsa.book.....O} appendix E, if we assume CI Cam is at a distance of 5kpc, our SED modelling seems in agreement with a B0-B1III classification. In order to be in agreement with the B0-B2 supergiant classification, the distance would have to be $\ge$8kpc. On the other hand, if we assume CI Cam is at the distance predicted by  \citet{2006ARep...50..664B} ie d=1.1-1.9~kpc, our SED modelling is in agreement with a B1-B2V spectral classification which is in agreement with  \citet{2006ARep...50..664B} Fig. 6 within the error.

Our best-fitting model to the interferometry data is a Gaussian skewed ring which confirms \citet{2002A&A...392..991H} and \citet{2002A&A...390..627M} predictions. The skewness of the ring can originate either from the formation of clumps during the phase of condensation of the dust or from the presence of a hot companion that would evaporate the dust in an non-homogeneous fashion. The ring is elliptical which confirms previous spectroscopy measurements \citep{2002A&A...390..627M,2007AJ....133.1478Y}. The semi-major axis matches the inner radius of the dust found for our DUSTY model (c) which is characterised by a lower T$_{eff}$, a larger dust grain population   and a somewhat lower luminosity with respect to previously proposed models. 

If we consider the close pair orbit in CI  Cam (B4III-V + WD) as defined in  \citet{2006ARep...50..664B} table~3 and assume that the orbit is coplanar with the dust ring, we can derive the semi-major axis of the orbit. From the Gaussian ring axis ratio, we find an inclination $i\approx67^{\circ}$. With $asin i =48\times10^{6}$~km, we derive the orbit semi-major axis $a=52\times10^{6}$~km=0.35~AU=74~R$_{\sun}$ which is well within the dust ring and unresolved with IOTA3T or PTI. We don't see any obvious sign of the third massive companion reported by \citet{2007ATel.1036....1B}. The search for a possible companion in the interferometry data being complicated by the extended asymmetric dust emission in CI Cam.

The ring like structure that we find, along with the very bright x-ray Fe K-line \citep{2000A&A...356..163O}, suggest that the luminosity emitted by the accreting object is reflected by the IR ring, giving rise to the Fe-K reflection spectrum. If this is indeed the source of the Fe-K line, then there should be a delay between
flux changes in the x-ray contiunuum and the Fe-K line, which should be equal to the light crossing time of 320 minutes if the size of the ring is indeed 40AU.  Detection of such a delay would allow an accurate geometric distance measure to the source by combining the x-ray reverberation delay and the IR interferometric angular size.  This would be the first such distance measure to an x-ray binary.

Future work should focus on the creation of a 3-D dust shell geometry model as well as self-consistent radiative transfer calculations. Additionally, further observations of CI Cam using the Michigan Infra-Red Combiner \citep{2006SPIE.6268E..55M} and CHARA Michigan phase-tracker \citep{2006SPIE.6268E.113B} at CHARA \citep{2005ApJ...628..453T} will be attempted in order to acquire model independent images. The high angular resolution and imaging capability would allow us to obtain an accurate image of the dust region morphology with the location and dimension of the asymmetries detected with IOTA3T.

\section*{Acknowledgements}
N. Thureau has received research funding from the European Community's Sixth Framework Programme through an International Outgoing Marie-Curie fellowship OIF - 002990.\\
We thank our IOTA colleagues for their invaluable contribution to the CI Cam observation program. The IONIC3 instrument has been developed by the Laboratoire d'Astrophysique de Grenoble (LAOG) and LETI in the context of the IONIC collaboration (LAOG, IMEP, LETI) with funding from the Centre National de Recherche Scientifique (CNRS, France) and Centre National d'Etudes Spatiales (CNES, France). \\
We acknowledge our PTI colleagues for their assistance in planning and carrying out the CI Cam observing program at PTI. PTI was developed by Jet Propulsion Laboratory and is operated by the Michelson Science Center on behalf of the PTI collaboration. \\
This research has made use of: NASA's Astrophysics Data System Bibliographic Services, SIMBAD database operated at CDS, Strasbourg, France, MSC resources,  CHARM2 and 2MASS catalogues through the VizieR service at CDS, Strasbourg, France.\\
TIFKAM was funded by the Ohio State University, the MDM consortium, MIT, and NSF grant AST-9605012. NOAO and USNO paid for the development of the ALADDIN arrays and contributed the array currently in use in    TIFKAM.\\
This research  was made possible  thanks to  a  Michelson Postdoctoral Fellowship  and  a Scottish  Universities  Physics Association  (SUPA) advanced fellowship  awarded to E.   Pedretti. \\
Part of the research described in this paper was carried out at the Jet Propulsion Laboratory, California Institute of Technology, under a contract with the National Aeronautics and Space Administration.\\
M.R. Garcia acknowledges partial support from NASA Contract NAS8-03060 to the Chandra X-ray Center.\\

\bibliography{cicam}
\bibliographystyle{mn2e}
\label{lastpage}
\end{document}